\documentclass[aps, prl, reprint, superscriptaddress]{revtex4-2}

\usepackage{graphicx}%
\usepackage{amsmath,amssymb,amsfonts}%
\usepackage{amsthm}%
\usepackage{xcolor}%
\usepackage{textcomp}%
\usepackage{times}
\usepackage{comment}
\usepackage{mathtools}
\usepackage{xspace}
\usepackage{color}

\begin{document}


\title{Chiral near-fields at achiral plasmonic tips enable nanoscale valley polarimetry}


\author{Jonggeun Hwang}
 \affiliation{Department of Physics, Pohang University of Science and Technology, Pohang 37673, Republic of Korea}
 \affiliation{Department of Physics, Massachusetts Institute of Technology, Cambridge, MA 02139, USA}
\author{Kijeong Park}
 \affiliation{Department of Physics, Pohang University of Science and Technology, Pohang 37673, Republic of Korea}
\author{Hyeongwoo Lee}
 \affiliation{Department of Physics, Pohang University of Science and Technology, Pohang 37673, Republic of Korea}
\author{Yeonjeong Koo}
 \affiliation{Department of Physics, Pohang University of Science and Technology, Pohang 37673, Republic of Korea}
 \affiliation{Department of Physics, Harvard University, Cambridge, MA 02138, USA}
\author{SeokJae Yoo}
  \affiliation{Department of Physics, Inha University, Incheon 22212, Republic of Korea}
\author{Long Ju}
 \affiliation{Department of Physics, Massachusetts Institute of Technology, Cambridge, MA 02139, USA}
\author{Kyoung-Duck Park}
 \email{Corresponding author: parklab@postech.ac.kr}
 \affiliation{Department of Physics, Pohang University of Science and Technology, Pohang 37673, Republic of Korea}
 \affiliation{Department of Semiconductor Engineering, Pohang University of Science and Technology, Pohang 37673, Republic of Korea}


\date{\today}

\begin{abstract}
Optical helicity is central to chiral light-matter interactions, yet plasmonic confinement typically distorts it at the nanoscale. 
Here, we demonstrate nanoscale valley polarimetry using an achiral Au tip, where incidence-angle control of circularly polarized light creates a strongly localized, helicity-preserving near-field. 
Chiral tip-enhanced photoluminescence (\textit{c}-TEPL) directly resolves and reverses local valley polarization in monolayer WSe$_2$ and reveals nanoscale polarization variations in strained MoS$_2$ hidden by diffraction-limited measurements.
\end{abstract}



\maketitle

\section{Introduction}

\noindent Valley polarization in atomically thin semiconductors is highly sensitive to local variations in strain, charge density, defects, and interfaces \cite{Schaibley2016valleytronics, Branny2017Deterministic, Moody2018Microsecond, Feng2019Engineering}.
In monolayer transition metal dichalcogenides (TMDs), these nanoscale perturbations reshape intervalley scattering, exciton relaxation, and transport, thereby controlling the degree of valley polarization (DOP) \cite{mak2012control, zeng2012valley}.
Yet, despite sustained efforts to control valley-polarized excitons, most measurements remain diffraction limited \cite{Schaibley2016valleytronics, Vitale2018Valleytronics, Yang2024Valleytronics}, leaving the nanoscale landscape that governs valley-selective optical responses largely unresolved. 
Nanoscale valley polarimetry capable of locally resolving the degree and sign of valley polarization is therefore needed to directly investigate these spatially varying processes.

Valley polarization can be engineered through strain \cite{Yang2023strain, Kumar2025strain}, doping \cite{Feng2019Engineering}, band-structure modification \cite{Liu2020room}, and heterostructure design \cite{Rivera2018Interlayer}.
In parallel, chiral optical fields have emerged as an effective route to strengthen or selectively drive circular-polarization-dependent optical transitions \cite{Sun2020Selectively, Wu2019room, Wen2020steering, Kim2020single}. 
However, a general physical principle for generating and controlling chiral near-fields at conventional achiral plasmonic tips has remained unclear.

Near-field nano-spectroscopy, including tip-enhanced photoluminescence (TEPL) and Raman scattering (TERS), provides the spatial resolution needed to access this regime \cite{Park2016hybrid, Xu2021Determining, Koo2021tip, Kato2022Ultrastable, Lee2024Quantum}, but plasmonic confinement introduces a fundamental trade-off, i.e., the near-field at a conventional metallic tip is dominated by strongly polarized out-of-plane fields that distort or suppress optical helicity \cite{Shao2019tip}.
Consequently, the helicity of incident circularly polarized light is weakened or distorted before it can efficiently drive valley-selective excitation. 
Chiroptical nanostructures \cite{Zhao2017nanoscopic} and spiral nanoprobes \cite{Long20233d} have been introduced to overcome this limitation.
More recently, efforts have been made to spatially resolve nanoscale valley polarization through a chiroptically selective multigrain apex structure \cite{Cheng2025visualizing}.
What is therefore needed is a simple and general way to generate and preserve chiral near-fields at conventional achiral plasmonic tips, thereby enabling nanoscale valley polarimetry without structurally chiral probes.

Here, we establish a general mechanism for generating chiral near-fields at achiral plasmonic tips, using the incidence angle of circularly polarized light as the control parameter. 
At an intermediate illumination angle, longitudinal plasmonic-mode excitation is balanced with preservation of the phase-shifted in-plane fields required for optical helicity. 
We demonstrate this principle through chiral tip-enhanced photoluminescence (\textit{c}-TEPL) in monolayer WSe$_2$ and establish it as a nanoscale valley polarimetry platform. The locally measured valley polarization reverses with excitation helicity while remaining negligible under linearly polarized excitation. We then apply this capability to strained MoS$_2$, where we uncover a previously inaccessible nanoscale valley-polarization landscape, revealing competing strain-induced depolarization and carrier-redistribution-dependent polarization enhancement that are obscured by diffraction-limited measurements.

\begin{figure*}[th!]
	\centering
	\includegraphics[width = \textwidth]{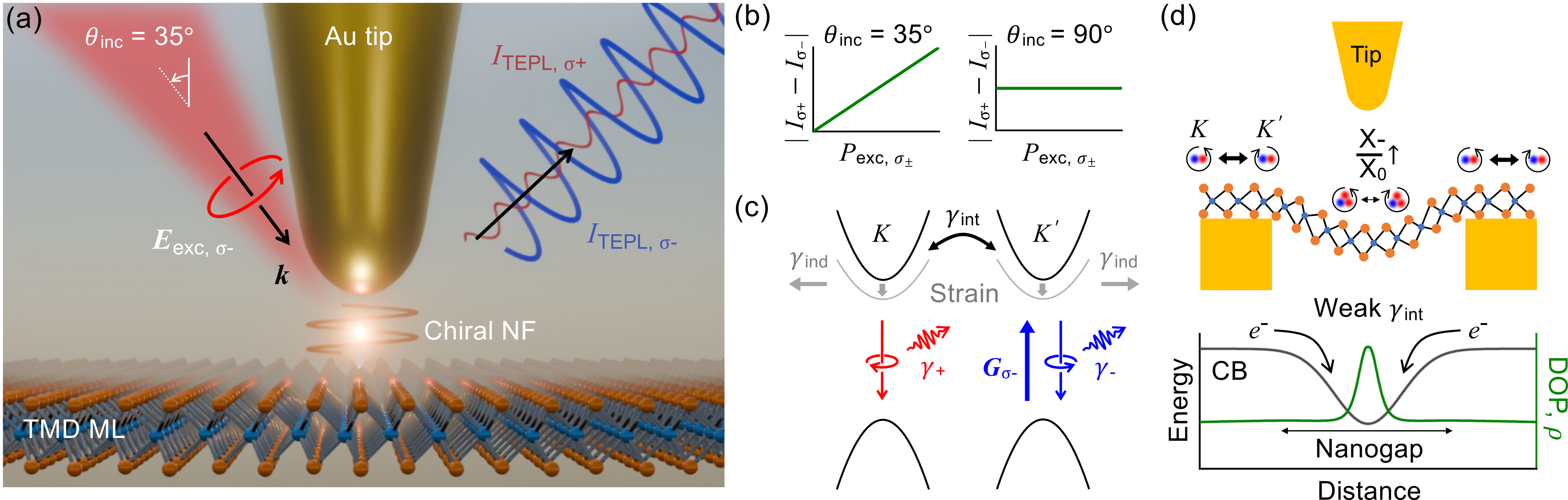}
	\caption{
		Illustration of \textit{c}-TEPL spectroscopy and nanoscale valley polarization heterogeneity.		
		(a) Schematic of \textit{c}-TEPL, in which oblique circularly polarized excitation generates a helicity-preserving near-field at an achiral Au tip for valley-resolved PL detection. 
		(b) Conceptual excitation-power dependence of the helicity-resolved PL contrast at $\theta_\text{inc} = 35^\circ$ and $90^\circ$.
		(c) Strain-induced modification of intervalley scattering pathways in a TMD monolayer. 
		$\gamma_\text{int}$ denotes the intervalley scattering rate, $\gamma_\text{ind}$ the indirect intervalley scattering rate, $G_{\sigma_-}$ the generation rate of the $K^\prime$ excitons, and $\gamma_+$ and $\gamma_-$ the radiative recombination rates of valley-polarized excitons. 
		(d) Schematic of valley-polarization redistribution across a suspended nanogap, where strain-driven carrier redistribution modifies the local trion population and DOP. CB, conduction band.
		}
\end{figure*}

\begin{figure*}[th!]
	\includegraphics[width= 1\textwidth]{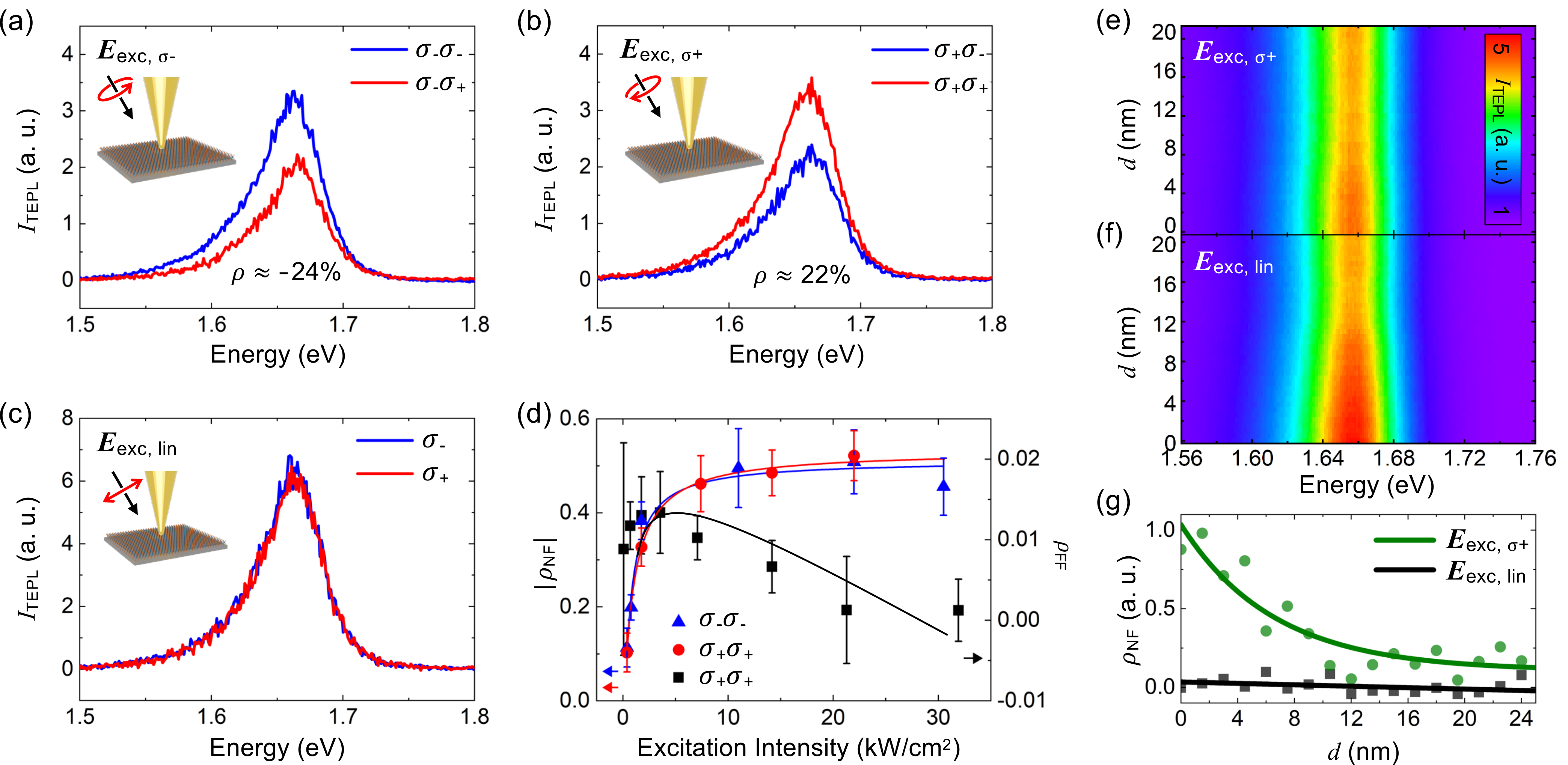}
	\centering
	\caption{
		Experimental demonstration of \textit{c}-TEPL-based nanoscale valley polarimetry.
		(a), (b) \textit{c}-TEPL spectra measured under $\sigma_-$ and $\sigma_+$ excitation, respectively. 
		(c) \textit{c}-TEPL spectra acquired under linearly polarized excitation.
		(d) Excitation intensity dependence of the near-field DOP magnitude (blue and red) and the far-field DOP (black). The error bars represent the standard deviation.
		(e), (f) Contour maps of gap-dependent TEPL spectra obtained under CPL and LPL excitation.
		(g) Gap-dependent DOP under two different excitation polarizations. 
		The DOP increases sharply for CPL excitation at tip-sample distances below 10 nm, whereas LPL excitation produces strong PL enhancement without a corresponding DOP increase.
		}
\end{figure*}
\section{\lowercase{\textit{c}}-TEPL-enabled nanoscale valley polarimetry in monolayer WS\lowercase{e}$_2$}

Figure 1(a) illustrates the \textit{c}-TEPL configuration, in which circularly polarized light is incident on a structurally achiral Au tip at $\theta_\text{inc} = 35^\circ$, while the emitted PL is resolved into the $\sigma_+$ and $\sigma_-$ channels. 
Oblique illumination generates a localized in-plane near-field with finite optical helicity, enabling valley-selective excitation. 
We first establish this helicity-resolved response in monolayer WSe$_2$ and then apply \textit{c}-TEPL to strained MoS$_2$ suspended across an Au nanogap, where strain and carrier redistribution produce spatially distinct modifications of valley polarization (Figs. 1(c)-1(d)).

We validate \textit{c}-TEPL using monolayer WSe$_2$ under $\sigma_+$, $\sigma_-$, and linearly polarized excitation. 
The near-field contribution is obtained by subtracting the far-field PL from the corresponding tip-coupled spectrum, and the DOP is calculated as
\begin{equation}
	\rho = \frac{I_{\sigma+} - I_{\sigma-}}{I_{\sigma+} + I_{\sigma-}},
\end{equation}
\noindent
where $I_{\sigma+}$ and $I_{\sigma-}$ are the near-field PL intensities detected in the two helicity channels.
The far-field DOP is nearly zero (see Supplemental Material, Fig. S2 \cite{Supplemental}\nocite{Yoo2018Polarimetric, Hwang2025compact, Ren2004preparation, harats2020dynamics, Rosati2021Dark, Mai2014many}), whereas the near-field spectra show pronounced helicity contrast. 
Under $\sigma_-$ and $\sigma_+$ excitation, the co-helicity emission is preferentially enhanced, yielding $\rho=-24\%$ and $+22\%$, respectively (Figs. 2(a)-2(b)). 
The comparable magnitudes and opposite signs demonstrate that the \textit{c}-TEPL response follows the excitation helicity rather than a fixed helicity preference of the tip-sample nanocavity.

Importantly, the observed chiral response does not originate from structural chirality of the probe, as the electrochemically etched Au tip is structurally achiral (Supplemental Material, Fig. S3 \cite{Supplemental}). 
Increasing the incidence angle from $35^\circ$ to $70^\circ$ substantially reduces the helicity-resolved contrast (Supplemental Material, Fig. S4 \cite{Supplemental}), demonstrating that the chiral response can be controlled through the illumination geometry without requiring optical chirality of the probe.

Polarization- and intensity-dependent measurements further distinguish helicity preservation from conventional tip enhancement. 
Under linearly polarized light (LPL) excitation, the $\sigma_-$ and $\sigma_+$ spectra nearly overlap despite strong PL enhancement, yielding negligible DOP (Fig. 2(c)) and ruling out a fixed helicity-selective emission bias. 
In the far field, the DOP peaks at an excitation intensity of $\mathord{\sim}5$ kW cm$^{-2}$ and decreases at higher intensities, consistent with enhanced intervalley depolarization and high-density excitonic processes \cite{Erkensten2021dark}. 
In contrast, the near-field $\vert\rho\vert$ remains comparatively robust with increasing intensity (Fig. 2(d)).
This behavior is consistent with an increased radiative recombination rate in the tip-enhanced near-field, allowing a larger fraction of excitons to emit while retaining their valley polarization before intervalley scattering occurs.
The nearly symmetric responses under opposite excitation helicities further exclude a fixed handedness of the probe.

Finally, gap-dependent measurements establish the near-field origin of the response. 
As the tip-sample distance decreases below $\mathord{\sim}10$ nm, LPL excitation produces strong PL enhancement without an appreciable DOP, whereas circularly polarized light (CPL) excitation produces a sharp increase in DOP over the same distance range (Figs. 2(e)-2(g)). 
Thus, \textit{c}-TEPL provides a localized helicity-resolved optical response rather than simply intensity-enhanced PL. 
The following numerical analysis identifies how oblique illumination generates and preserves this chiral near-field.
\begin{figure*}[th!]
	\centering
	\includegraphics[width= 0.85\textwidth]{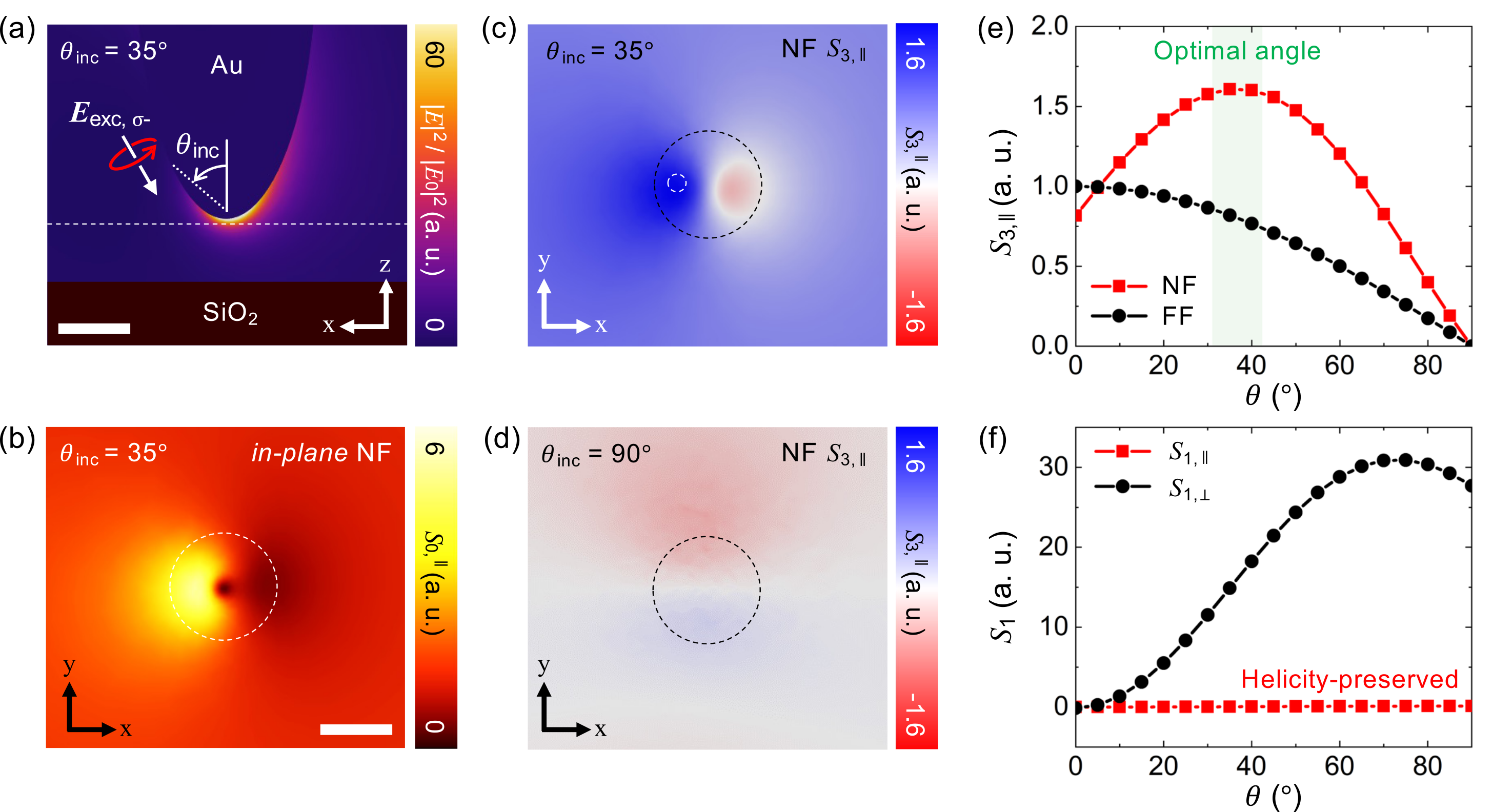}
	\caption{
		Numerical simulation results for \textit{c}-TEPL as a function of incidence angle.
		(a) Normalized near-field intensity, $\vert{E}\vert^2/\vert{E_0}\vert^2$, in the $xz$-plane for $\theta_\text{inc} = 35^\circ$.
		(b), (c) In-plane Stokes parameters $S_{0,\parallel}$ and $S_{3,\parallel}$ in the $xy$ plane, respectively, evaluated 1 nm below the tip apex. 
		(d) Spatial distributions of the Stokes parameter $S_{3,\parallel}$ in the same plane as (c) for  $\theta_\text{inc} = 90^\circ$. The dashed circle has a radius equal to the tip’s radius of curvature. All scale bars are 20 nm.
		(e) $S_{3,\parallel}$ for the near-field and far-field in the $xy$-plane as a function of $\theta_\text{inc}$. The plotted near-field $S_{3,\parallel}$ values are taken at the maximum-$S_{3,\parallel}$ position (small white circle in (c)).
		(f) $\theta_\text{inc}$-dependent in-plane and out-of-plane linear-polarization contrast, $S_{1,\parallel}$ and $S_{1,\bot}$.
		}
\end{figure*}

\section{Numerical analysis of near-field helicity transfer in \lowercase{\textit{c}}-TEPL}

To identify the electromagnetic mechanism underlying the experimentally observed angle dependence of \textit{c}-TEPL, we perform full-wave simulations using COMSOL Multiphysics while varying the incidence angle, $\theta_\text{inc}$.
Because the valley-selective optical transitions of monolayer WSe$_2$ couple to in-plane circular dipoles, \textit{c}-TEPL requires not only a strong local field but also a helicity-preserving in-plane near-field at the sample plane.
We therefore analyze the $x$- and $y$-field components using the in-plane Stokes parameters \cite{berry1977measurement}.
\begin{eqnarray}
	S_{0, \parallel} = \vert E_x \vert^2 + \vert E_y \vert^2 \\
	S_{1, \parallel} = \vert E_x \vert^2 - \vert E_y \vert^2 \\
	S_{2, \parallel} = 2\text{Re}(E_x E_y^*) \\
	S_{3, \parallel} = -2\text{Im}(E_x E_y^*)
\end{eqnarray}
\noindent Here, $S_{0,\parallel}$ represents the in-plane field intensity, $S_{1,\parallel}$ and $S_{2,\parallel}$ the linear polarization components, and $S_{3,\parallel}$ the circular polarization component. 
Under the valley optical selection rules, the excitation rates satisfy $W_\pm\propto(S_{0,\parallel}\pm S_{3,\parallel})/2$. 
A strong \textit{c}-TEPL response therefore requires spatial overlap between a large $S_{0,\parallel}$ and a finite $S_{3,\parallel}$.

At $\theta_\text{inc}=35^\circ$, the enhanced field is strongly localized at the tip apex (Fig. 3(a)), where a large $S_{0,\parallel}$ overlaps with a finite $S_{3,\parallel}$ 1 nm below the apex (Figs. 3(b)-3(c)). 
At normal incidence ($\theta_\text{inc}=0^\circ$), inefficient coupling to the longitudinal tip mode leaves both components weak beneath the apex (see Supplemental Material, Figs. S5-S10 \cite{Supplemental} for the full results including the results of $\sigma_-$ excitation). 
Increasing $\theta_\text{inc}$ strengthens longitudinal-mode excitation while retaining the transverse field component required for finite $S_{3,\parallel}$. 
At $\theta_\text{inc}=35^\circ$, these conditions coincide, producing strong in-plane confinement while preserving the incident helicity. 
Small regions of opposite $S_{3,\parallel}$ occur away from the field maximum, but their weak $S_{0,\parallel}$ makes their contribution to valley-selective excitation negligible. 
As $\theta_\text{inc}$ approaches $90^\circ$, the balance between the two in-plane components is lost and $S_{3,\parallel}$ nearly vanishes beneath the apex (Fig. 3(d)).

The angle dependence makes this trade-off explicit, i.e., the near-field $S_{3,\parallel}$ peaks near $35^\circ$ and decreases toward both smaller and larger angles (Fig. 3(e)). 
This optimum reflects a balance between efficient excitation of the longitudinal plasmonic mode and preservation of the phase-shifted in-plane field components required for circular polarization. 
Because this balance depends on tip geometry, the precise optimal angle is expected to vary with apex curvature and aspect ratio. 
Consistently, near the optimal angle the in-plane linear Stokes components $S_{1,\parallel}$ and $S_{2,\parallel}$ remain small while $S_{3,\parallel}$ remains finite (see Fig. 3(f) and Supplemental Material, Fig. S11 \cite{Supplemental}), consistent with a predominantly circular in-plane near-field.

\section{Nanoscale valley-polarization redistribution revealed by \lowercase{\textit{c}}-TEPL}

Having established the helicity-resolved near-field response of \textit{c}-TEPL, we use it to investigate how nanoscale strain and carrier redistribution reshape valley polarization. 
An intrinsically electron-rich monolayer MoS$_2$ \cite{Singh2019Origin} is suspended across a 120-nm-wide Au nanogap (Supplemental Material, Fig. S12 \cite{Supplemental}), producing strong spatial variations in strain and carrier density. 
This geometry therefore provides a platform to disentangle their respective effects on valley polarization.

Conventional TEPL characterization (see Supplemental Material, Fig. S13 \cite{Supplemental}) reveals an enhanced $X$-$/X_0$ ratio toward the nanogap center (Fig. 4(d)) and a 3-4 meV PL redshift within the suspended region (Supplemental Material, Fig. S14 \cite{Supplemental}), corresponding to $\mathord{\sim}0.1\%$ tensile strain \cite{Schmidt2016Reversible}. 
These signatures indicate spatially varying strain and electron density across the nanogap \cite{Lee2022drift, Lee2023all}, but do not reveal how these inhomogeneities reshape the local valley polarization.

We therefore acquire helicity-resolved \textit{c}-TEPL images under fixed $\sigma_+$ excitation and calculate the local DOP pixel by pixel (Figs. 4(a)-4(c)). 
The resulting DOP map reveals a pronounced W-shaped profile across the nanogap, i.e., the polarization is suppressed near both boundaries and enhances toward the center (Fig. 4(g)). 
This nanoscale structure is absent from the diffraction-limited confocal DOP image of the same region (Fig. 4(h) and Supplemental Material, Fig. S15 \cite{Supplemental}), where the distinct edge and center responses are spatially averaged.

The spatial correlations suggest two distinct mechanisms underlying this profile. The DOP minima occur near the nanogap boundaries, where the strain gradient is largest, consistent with strain-enhanced intervalley depolarization \cite{Kumar2025strain}. 
In contrast, the enhancement of DOP toward the center coincides with an enhanced $X$-$/X_0$ ratio (Figs. 4(d), 4(e), and 4(g)), indicating increased trion population from strain-driven electron accumulation. 
Because trions are less susceptible to exchange-mediated valley depolarization than neutral excitons \cite{Wang2018Colloquium}, this carrier redistribution provides a mechanism for stabilizing valley polarization near the nanogap center.

To examine the contribution of strain-driven exciton transport, we solve coupled valley-resolved drift-diffusion equations for the two valley populations, incorporating the strain-induced potential, exciton transport, recombination, and intervalley relaxation (Supplemental Material, Supplemental Note and Fig. S16 \cite{Supplemental}). 
The calculated valley-population imbalance, $\Delta n_v(x)=n_+(x)-n_-(x)$, reproduces the overall spatial redistribution observed experimentally (Fig. 4(i)), supporting a role for strain-driven exciton transport. 
The additional DOP suppression near the nanogap boundaries is consistent with enhanced intervalley depolarization in regions of large strain gradient.
\begin{figure*}[th!]
	\centering
	\includegraphics[width = 1\textwidth]{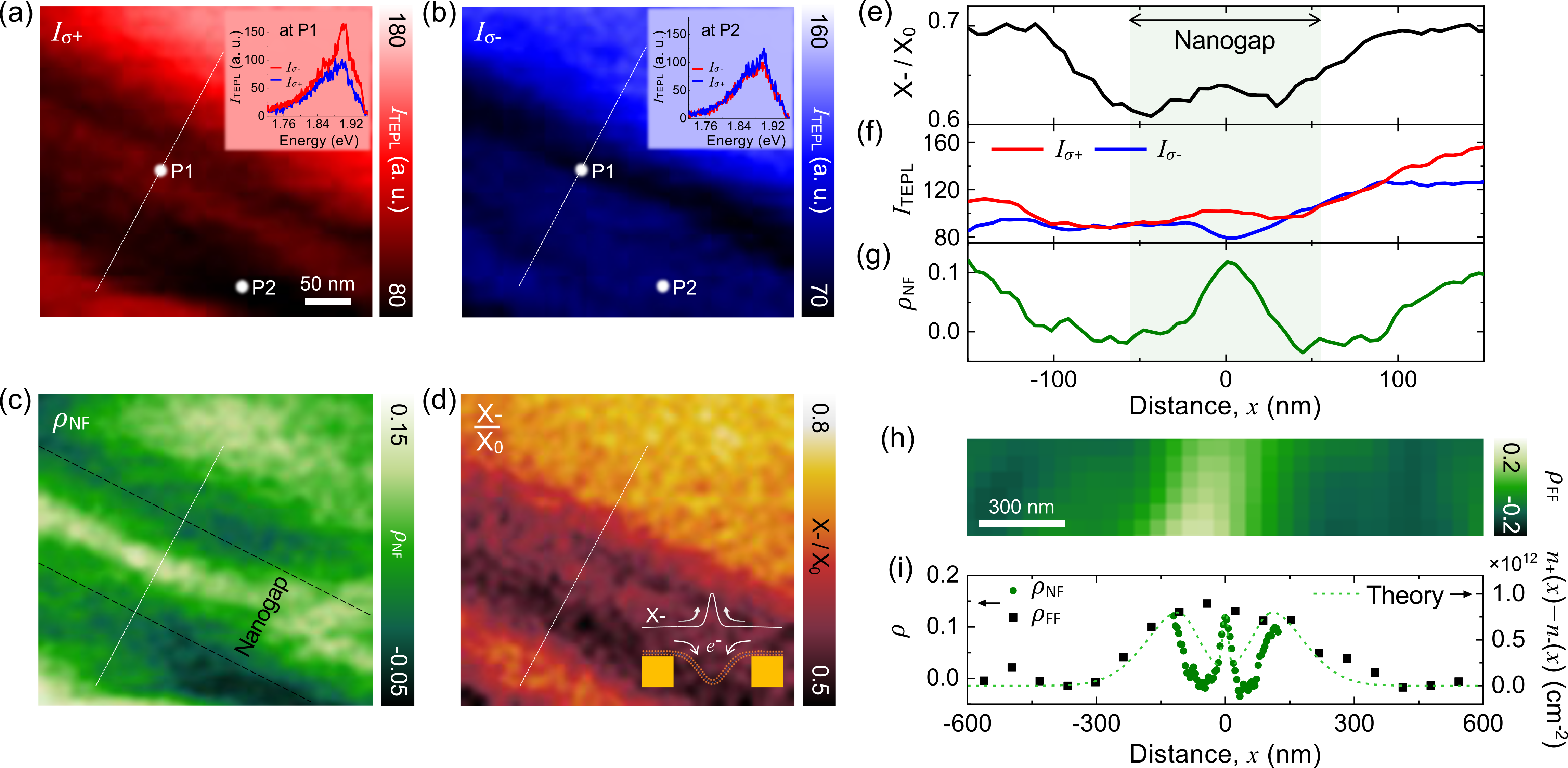}
	\caption{
	    	Nanoscale DOP imaging of a suspended TMD monolayer across a nanogap.	
		(a), (b) \textit{c}-TEPL images of the $\sigma_+$ PL and $\sigma_-$ PL. The insets show spectra acquired at the marked positions.
		(c) Nanoscale DOP image, $\rho_\text{NF}$, obtained from (a) and (b). The black dashed lines indicate the boundaries of the nanogap. 
		(d) Trion-to-exciton ratio ($X$-$/ X_0$) map, obtained from Lorentzian fits to the TEPL spectra.	
		(e)-(g) Line profiles across the nanogap showing $X$-$/ X_0$, $\sigma_+$ and $\sigma_-$ PL intensities, and DOP, respectively. The shaded area marks the nanogap region.
		(h) Far-field DOP map of the nanogap region measured by confocal PL microscopy.
		(i) Comparison of far-field DOP (black squares) and near-field DOP (green circles) with the theoretically predicted spatial distribution of $n_+(x) - n_-(x)$ (green dotted line).
		}
\end{figure*}
\section{Conclusions}

In conclusion, we establish a simple mechanism for generating helicity-preserving near-fields at structurally achiral plasmonic tips by controlling the incidence angle of circularly polarized light. 
This enables \textit{c}-TEPL without engineered chiral probes and provides nanoscale access to local valley polarization beyond the diffraction limit. 
Applied to strained MoS$_2$, \textit{c}-TEPL reveals a W-shaped DOP profile associated with strain-enhanced depolarization near the nanogap boundaries and polarization stabilization toward the center. 
Valley-resolved drift-diffusion calculations further support a contribution from strain-driven exciton redistribution. 
More broadly, our results establish optical helicity as a controllable degree of freedom in tip-enhanced nano-spectroscopy. 
Because the chiral near-field is generated through illumination geometry rather than structural chirality of the probe, this principle should be adaptable to other near-field spectroscopic modalities and material systems. 
It could enable nanoscale interrogation of spin- and valley-selective excitations \cite{Xu2014Spin}, chiral phonons \cite{Zhu2018Observation}, molecular chirality \cite{Tang2011Enhanced}, and other helicity-dependent light-matter interactions that remain inaccessible to diffraction-limited spectroscopy. 
Our approach thus transforms a conventional achiral plasmonic tip into a general nanoscale probe of chiral light–matter interactions.

\vspace{\baselineskip}
\begin{acknowledgments}
This work was supported by National Research Foundation of Korea (NRF) grants (RS-2025-02217103, RS-2025-00559639, RS-2025-17492968, 2021R1A6A1A1004294413, RS-2025-02317602, and RS-2022-NR068228), the Samsung Science and Technology Foundation (SSTF-BA2102-05) and the Ministry of Science and ICT (MSIT) under the Information Technology Research Center (ITRC) support program (IITP-2022-RS-2022-00164799). 
\end{acknowledgments}
\bibliography{references}

\end{document}